\documentclass[twocolumn,showpacs,preprintnumbers]{revtex4}

\usepackage{amsmath}
\usepackage{graphicx}
\usepackage{dcolumn}
\usepackage{bm}
\usepackage{color}

\begin{document}

\title{Single-photon scattering controlled by an imperfect cavity}

\author{Liwei Duan$^{1}$}
\author{Qing-Hu Chen$^{1,2}$}\email{qhchen@zju.edu.cn}

\affiliation{
	$^{1}$ Department of Physics and Zhejiang Province Key Laboratory of Quantum Technology and Device, Zhejiang University, Hangzhou 310027, China \\
	$^{2}$ Collaborative Innovation Center of Advanced Microstructures,  Nanjing University,  Nanjing 210093, China}
\date{\today }

\begin{abstract}
	We study the  single-photon transport in the coupled-resonator waveguide (CRW) controlled by an imperfect cavity. A Lorentzian spectrum is introduced to describe the dissipation. We find that the probability current  conservation can be broken, although the imperfect cavity is a Hermitian system. The coupling strength between the imperfect cavity and the CRW has significant influences  near the resonant frequency. With the increase of the coupling strength, the transmission coefficient becomes smaller. The spectral width plays a dominant role under the off-resonant condition, where the transmission  coefficient is greatly suppressed with the increase of spectral width.
	We also observe an abrupt jump of the transmission and reflection coefficients, when the hopping amplitude is large enough.
	All the distinctive  behaviors are closely related to the complex effective potential  induced by the imperfect cavity.
\end{abstract}


\maketitle

\section{Introduction}

Quantum optical switches are quantum systems to control the flow of light \cite{PhysRevLett.71.2360,orrit2007quantum,Chen768, PhysRevLett.111.193601}, which play a significant role in the quantum circuits, quantum communication and quantum networks \cite{PhysRevLett.78.3221, kimble2008quantum, RevModPhys.82.1209, ritter2012elementary}. One of the most notable quantum optical  {switches} is based on the controllable scattering of a single photon in low-dimensional waveguides \cite{PhysRevLett.95.213001,PhysRevLett.101.100501}. The waveguide transfers the photon which carries  information, while the scattering center coupled to the waveguide serves as a switch to control the transmission and reflection of the photon.

Early studies mainly focus on the waveguide which is an one-dimensional continuum with linear dispersion \cite{doi:10.1063/1.1448174,PhysRevE.62.7389, Shen:05,PhysRevLett.95.213001,PhysRevA.79.023837,PhysRevA.79.023838,TIANWei:30302}. Shen and Fan proposed a real-space Hamiltonian with which they predicted many novel properties of single photon transport, such as the general Fano line shape and symmetric vacuum Rabi splitting for a leaky cavity\cite{PhysRevLett.95.213001}. Recently, {an} one-dimensional coupled-resonator waveguide (CRW) with cosine-type dispersion gets more and more attention, which can be realized in the experiment by using superconducting transmission line resonators \cite{PhysRevLett.101.100501}. Due to the nonlinear dispersion relation, the reflection amplitude shows a more general structure, beyond the Breit-Wigner and Fano line shapes \cite{PhysRevLett.101.100501}. Two CRWs have been combined to study the quantum routing of single photon controlled by a cyclic three-level system \cite{PhysRevLett.111.103604}. Most recently, a dimerized CRW was proposed to study the scattering process which acts as a photonic analog of the Su-Schrieffer-Heeger model \cite{Belloeaaw0297,zhang2019transmission}. The dimerized CRW leads to unconventional scattering properties and different super- and sub-radiant states depending on the band topology \cite{Belloeaaw0297}. Despite of the differences, all the waveguides mentioned above are Hermitian systems. The {non-Hermitian} waveguides with $\mathcal{PT}$-symmetry were constructed \cite{PhysRevLett.106.213901,PhysRevA.85.023802,doi:10.1002/andp.201900120}, which lead to some singular optical characteristics, such as unidirectional invisibility and anisotropic transmission.

The scattering centers have much more choices than the waveguides. A single-mode cavity or resonator is one of the simplest scattering centers which can be direct or side coupled to the waveguide \cite{doi:10.1063/1.1448174,PhysRevE.62.7389}. Besides, the interaction between them can be linear \cite{PhysRevA.79.023837} or nonlinear \cite{Zhou:20}. Recent experiments have demonstrated that the light-matter interaction can reach {ultrastrong} and even deep-strong coupling regimes \cite{RevModPhys.91.025005,braak2016semi,Shen:05}, which make the multi-level atom or superconducting qubit a good candidate to be the scattering center \cite{PhysRevLett.71.2360,PhysRevLett.95.213001,PhysRevLett.101.100501,Zhang:13,GULing-Ming:104206}. Additionally,  {a} two-level atom can be {embedded} in a cavity, while the cavity is coupled to the waveguide \cite{PhysRevA.79.023837,PhysRevA.79.023838}.
The hybrid atom-optomechanical systems can also be regarded as the scattering center, which reveal richer spectral features \cite{PhysRevA.87.033807,PhysRevA.88.063821,PhysRevA.96.013860,Yan:15}.

Due to the influence of the environment, it is hard to construct a perfect single-mode cavity with no dissipation \cite{breuer2002theory}. In the previous studies, one usually introduces an imaginary value to the  cavity frequency to account for the dissipation \cite{PhysRevA.79.023837,PhysRevA.79.023838,Witthaut_2010,PhysRevA.88.063821,Yan:15}, which forms a non-Hermitian scattering center. For the non-Hermitian scattering center, the probability current is usually not conserved. However, a $\mathcal{PT}$-symmetric scattering center can preserve the probability current conservation \cite{PhysRevA.85.012111} and change the perfect reflection to perfect transmission at the resonant frequency \cite{PhysRevA.91.042131}. It should be noted that the introduction of the imaginary cavity frequency to describe the environment-induced dissipation is not exact \cite{PhysRevLett.68.580,PhysRevLett.70.2273,PhysRevLett.123.170401}. In this paper, we introduce an imperfect cavity to play the role of scattering center. The unavoidable cavity dissipation is taken into account by assuming a Lorentzian spectrum for the cavity \cite{breuer2002theory,Cao_2011,PhysRevLett.99.160502,Wang_2013,esfandiarpour2019cavity}, rather than an imaginary frequency. The imperfect cavity is coupled with one resonator of the CRW. A single incident photon transferring in the CRW can  be scattered by the imperfect cavity. The paper is organized as follows. In section II, the Hamiltonians of the CRW and the imperfect cavity are introduced. The Hamiltonian of the perfect cavity is also listed for comparison. In section III, we briefly introduce the method to deal with the scattering process and derive the transmission and reflection coefficients. In section IV, we compare the results of the perfect and imperfect {cavities}. The influences of the imperfect cavity are analyzed in details.  The last section contains some concluding remarks.

\section{Hamiltonian}

\begin{figure}[htbp]
	\centering
	\includegraphics[width=8cm]{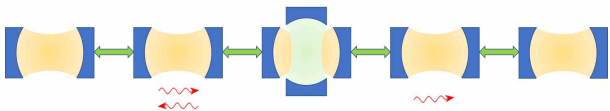}
	\caption{(color online) Schematic configuration for the coherent transport of a single
		photon in a CRW. A  cavity is coupled to the $0$-th resonator,
		which controls the transport of the incident photon.}
	\label{CRW1}
\end{figure}

As shown in Fig. \ref{CRW1}, we consider a 1D CRW described by
\begin{eqnarray}
H_{\mathrm{c}}=\Omega \sum_{j}a_{j}^{\dag }a_{j}-J\sum_{j}\left( a_{j}^{\dag
}a_{j+1}+a_{j+1}^{\dag }a_{j}\right) ,  \label{Hc}
\end{eqnarray}
where $a_j$ ($j=-\infty, \dots, +\infty$) is the annihilation operator of
the $j$-th resonator with frequency $\Omega$, and $J$ is the intercavity
hopping amplitude. The Hamiltonian (\ref%
{Hc}) is a typical tight-binding boson model which has the dispersion
\begin{eqnarray}
E_{k}=\Omega -2J\cos k.
\end{eqnarray}

The $0$-th resonator is coupled with an {additional} cavity  {which serves as the scattering center. Here we consider two kinds of cavities: a perfect single-mode cavity and an imperfect cavity which introduces the influence of the environment-induced dissipation.}

\subsection{ {Perfect cavity}}

 {If the $0$-th resonator of the 1D CRW is coupled to a perfect single-mode cavity, then} the Hamiltonian under the rotating wave approximation can be written as
\begin{eqnarray}
H_{\mathrm{I}}^{(\mathrm{p})}=\omega _c b^{\dag }b + \eta \left( b^{\dag
}a_{0}+a_{0}^{\dag }b\right) , \label{perfect_cavity}
\end{eqnarray}
where $b$ ($b^{\dagger}$) is the annihilation (creation) operator of the single mode, $\omega_{\mathrm{c}}$ is the frequency of the cavity, and $\eta$ is the coupling strength between the cavity and the $0$-th resonator. The total Hamiltonian related to the perfect cavity  can be written as
\begin{eqnarray}
H^{(\mathrm{p})}=H_{\mathrm{I}}^{(\mathrm{p})}+H_{\mathrm{c}} ,
\end{eqnarray}
The influence of the perfect cavity  has been widely studied \cite{doi:10.1063/1.1448174,PhysRevE.62.7389,PhysRevLett.101.100501,Zhou:20}. At the resonant frequency $E_k=\omega_{\mathrm{c}}$, the perfect cavity leads to total reflection.

\subsection{ {Imperfect cavity}}

 {If the $0$-th resonator of the 1D CRW is coupled to an imperfect cavity, then} the corresponding Hamiltonian \cite{breuer2002theory,Cao_2011,PhysRevLett.99.160502,Wang_2013,esfandiarpour2019cavity} would be
\begin{eqnarray}
H_{\mathrm{I}}^{(\mathrm{i})} = \sum_{q}\omega _{q}b_{q}^{\dag }b_{q}+ \sum_{q}g_{q}\left( b_{q}^{\dag
}a_{0}+a_{0}^{\dag }b_{q}\right) ,
\end{eqnarray}
where $b_q$ ($b_q^{\dagger}$) is the annihilation (creation) operator of the $q$-th mode with frequency $%
\omega_q$. For the imperfect cavity, due to the environment-induced cavity leakage, it can be described by a Lorentzian spectrum $J(\omega)$ with a central frequency, namely a single-mode cavity with its frequency broadened by the cavity leakage. The interaction strength $g_q$ is determined by the spectral
density $J(\omega)$, which can be written as
\begin{eqnarray}
J(\omega)&=&\sum_{q} g_q^2 \delta(\omega - \omega_{q})  \notag \\
&=& \frac{1}{2 \pi} \frac{\lambda \gamma^2}{\left( \omega_{\mathrm{c}} -
	\omega\right)^2 + \gamma^2}, \label{spectra_density}
\end{eqnarray}
where $\omega_{\mathrm{c}}$ is the central frequency of the cavity, $\lambda$ is the coupling strength which determines the maximum of the spectral density, and $\gamma$ defines the spectral width of the coupling which is connected to the dissipation rate of the system.
The total Hamiltonian related to the imperfect cavity can be written as
\begin{eqnarray}
H^{(\mathrm{i})}=H_{\mathrm{I}}^{(\mathrm{i})}+H_{\mathrm{c}} ,
\end{eqnarray}
which has a $U(1)$ symmetry. It commutes with the total excitation number
operator $N=\sum_{j}a_{j}^{\dagger }a_{j} + \sum_{q}b_{q}^{\dagger }b_{q}$,
namely $[H^{(\mathrm{i})},N]=0$.

\section{Scattering process}

Here we focus on the single-photon elastic scattering. The stationary
eigenstate corresponding to momentum $k$ is assumed to be
\begin{equation*}
|\phi _{k}\rangle =\sum_{j}\alpha _{j}a_{j}^{\dag }|0\rangle +\sum_{q}\beta
_{q}b_{q}^{\dag }|0\rangle .
\end{equation*}%
where $|0\rangle $ is the vacuum state of all the resonators and cavity.
From the Schr\"{o}dinger equation $H^{(\mathrm{i})}|\phi _{k}\rangle = {\color{red}{E_k}} |\phi
_{k}\rangle $, one can easily get that
\begin{eqnarray}
E_{k}\beta _{q} &=&\omega _{q}\beta _{q}+g_{q}\alpha _{0},  \label{EOM1} \\
E_{k}\alpha _{j} &=&\Omega \alpha _{j}-J\left( \alpha _{j+1}+\alpha
_{j-1}\right) +\delta _{j,0}\sum_{q}g_{q}\beta _{q}.  \label{EOM2}
\end{eqnarray}
Combining  Eqs. (\ref{EOM1}) and (\ref{EOM2}), we can achieve the following
discrete scattering equation,
\begin{equation}
\left( \Omega -E_{k}+V \delta _{j,0}\right) \alpha _{j}=J\left( \alpha
_{j+1}+\alpha _{j-1}\right) , \label{scattering_eq}
\end{equation}%
where
\begin{widetext}
\begin{eqnarray}
V &=&\sum_{q}\frac{g_{q}^{2}}{E_{k}-\omega _{q}} =\int_{0}^{+\infty }d\omega \frac{J\left( \omega \right) }{E_{k}-\omega } \label{potential}\\
&=&\frac{\lambda \gamma \left[ \gamma \mathrm{Re}\left( \ln \left( -\frac{1}{%
		\omega_{\mathrm{c}}-\mathrm{i} \gamma }\right) \right)-\left( E_{k}-\omega_{\mathrm{c}}\right) \mathrm{Im}%
	\left( \ln \left( -\frac{1}{\omega_{\mathrm{c}}-\mathrm{i} \gamma }\right) \right) -\gamma \ln
	\left( -\frac{1}{E_{k}}\right) \right] }{2 \pi \left( \left(
	E_{k}-\omega_{\mathrm{c}}\right) ^{2}+\gamma ^{2}\right) } .  \notag
\end{eqnarray}
\end{widetext}
is the effective scattering potential at site $j=0$. Unlike the effective potential induced by a two-level system or  perfect single-mode cavity which is a real number \cite{PhysRevLett.101.100501,Zhou:20}, $V$ is a complex one here. The cavity mode with $\omega=E_k$ (or $\omega_{q}=E_k$) plays a dominant role on the scattering process, since the incident photon resonates with this cavity mode. As the range of the cavity frequency  $\omega$ is $\left(0, +\infty\right)$, the incident photon with energy $E_k$ will always resonate with one of the modes as long as $E_k>0$. From Eq. (\ref{potential}), one can easily verify that the imaginary part of the effective potential $V$ is
\begin{eqnarray}
\mathrm{Im} \left(V\right) = \left\{
\begin{array}{ll}-\frac{\lambda \gamma^2  }{2 \left( \left(	E_{k}-\omega_{\mathrm{c}}\right) ^{2}+\gamma ^{2}\right) } & \mathrm{,~if}~ E_k>0,\\
0 & \mathrm{,~if}~ E_k<0.
\end{array}%
\right. \label{Im_V}
\end{eqnarray}
Therefore, one can predict that the imperfect cavity has more  {influences} on the incident photon with energy $E_k>0$.

A photon with momentum $k \in \left(0, \pi\right)$ incident from the left side of the CRW will
result in transmitted and reflected photon. For $j\neq 0$,%
\begin{equation*}
\alpha _{j}=\left\{
\begin{array}{ll}
\mathrm{e}^{\mathrm{i}kj}+r\mathrm{e}^{-\mathrm{i}kj} & \mathrm{,~if}~j<0, \\
t\mathrm{e}^{\mathrm{i}kj} & \mathrm{,~if}~j> 0,%
\end{array}%
\right.
\end{equation*}%
where $t$ and $r$ are the transmission and reflection amplitudes respectively. The continuous condition $\alpha_{j=0^+}=\alpha_{j=0^-}$ leads to $1+r=t$. From the discrete scattering equation (\ref{scattering_eq}) at $j=0$, together with the continuous condition,
we can obtain the transmission and reflection amplitudes
\begin{eqnarray}
t &=&\frac{2 \mathrm{i}J\sin k}{2 \mathrm{i}J\sin k-V}, \\
r &=&\frac{V}{2 \mathrm{i}J\sin k-V},
\end{eqnarray}%
with which the transmission coefficient $T$ and the reflection coefficient $R$ are achieved accordingly
\begin{eqnarray}
T(k)&=&\left\vert t\left( k\right) \right\vert ^{2} =\frac{4J^{2}\sin ^{2}k}{%
	\mathrm{Re}\left( V\right) ^{2}+\left( 2J\sin k-\mathrm{Im}\left( V\right)
	\right) ^{2}}, \label{T_k}\\
R(k)&=&\left\vert r\left( k\right) \right\vert ^{2} =\frac{\mathrm{Re}\left(
	V\right) ^{2}+\mathrm{Im}\left( V\right) ^{2}}{\mathrm{Re}\left( V\right)
	^{2}+\left( 2J\sin k-\mathrm{Im}\left( V\right) \right) ^{2}}. \label{R_k}
\end{eqnarray}

\section{Results and discussions}

The imperfect cavity introduces the influence of dissipation on the perfect cavity. We expect that the imperfect cavity can recover the perfect one in the limit of weak dissipation $\gamma \rightarrow 0$. According to the definition of the Dirac delta function,
\begin{equation*}
\delta \left( x-x_{0}\right) =\lim_{a\rightarrow 0^{+}}\frac{1}{\pi }\frac{a%
}{\left( x-x_{0}\right) ^{2}+a^{2}},
\end{equation*}
at $\gamma \rightarrow 0$ one can rewrite the spectral density (\ref{spectra_density}) as
\begin{eqnarray}
J(\omega ) = \sum_{q} g_q^2 \delta(\omega - \omega_{q})  \simeq \frac{\lambda \gamma }{2}\delta \left( \omega -\omega _{c}\right).
\end{eqnarray}%
Only one mode $q=\bar{q}$ exists which satisfies $\omega _{q=\bar{q}}=\omega _{c}$ and $%
g_{q=\bar{q}}=\sqrt{\lambda \gamma / 2}$, while other modes have no
influences since $g_{q\neq \bar{q}}=0$. This system is equivalent to the perfect cavity (\ref{perfect_cavity}) with frequency $\omega_{\mathrm{c}}$ and coupling strength $\eta = \sqrt{\lambda \gamma / 2}$.

\begin{figure}[htbp]
	\centering
	\includegraphics[width=9cm]{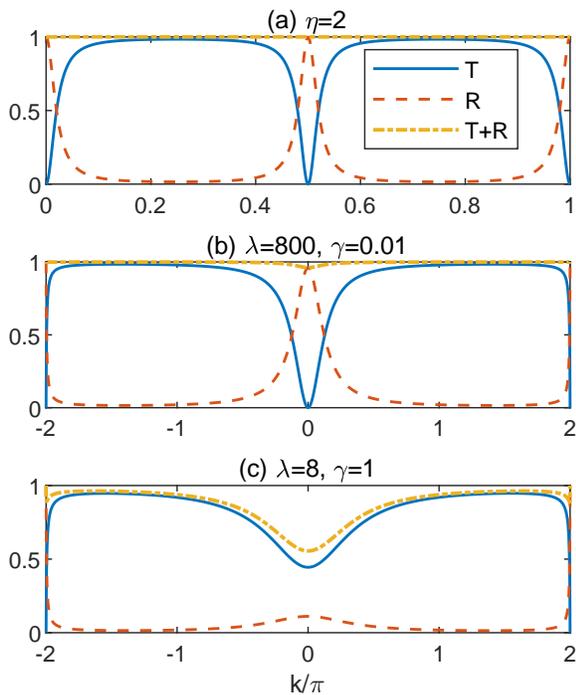}
	\caption{(color online) The transmission coefficient $T$ (blue solid line),  reflection coefficient $R$ (red dashed line)  {and their summation $T + R$ (yellow dash-dot line)} as a function of momentum $k$ at $\Omega=\omega_{\mathrm{c}}=10$, $J=4$. For the perfect cavity, (a) $\eta=0.1$. For the imperfect cavity, (b) $\lambda=800$, $\gamma=0.01$ and (c) $\lambda=8$, $\gamma=1$.}
	\label{TR_k}
\end{figure}

Fig. \ref{TR_k} (a) depicts the transmission and reflection coefficients of the perfect cavity as a function of $k$.   {It is obvious that $T+R=1$ as indicated by the yellow dash-dot line.} The vanishing transmission coefficients at $k=0$ and $ \pi$ are trivial, which are due to the vanishing group velocity at those points for the cosine-type dispersion relation. However, the vanishing transmission coefficient at $k=\pi /2$ is due to the effective potential induced by the perfect cavity at the resonance frequency $E_k=\omega_{\mathrm{c}}$ \cite{PhysRevLett.101.100501}. Figs. \ref{TR_k} (b) and (c) give the transmission and reflection coefficients of the  imperfect cavities. Here we have made sure that $\sqrt{\lambda \gamma / 2}$ of the imperfect cavity equals to the coupling strength $\eta$ of the perfect cavity. In the limit of $\gamma \rightarrow 0$, the imperfect cavity is equivalent to the perfect one as  shown in Fig. \ref{TR_k} (b). When we further increase $\gamma$, the distinctive behaviors of the imperfect cavity emerge. It is noteworthy that  $(R+T)$ may not be one,  {as shown in Fig. \ref{TR_k} (c)}.

\begin{figure}[htbp]
	\centering
	\includegraphics[width=9cm]{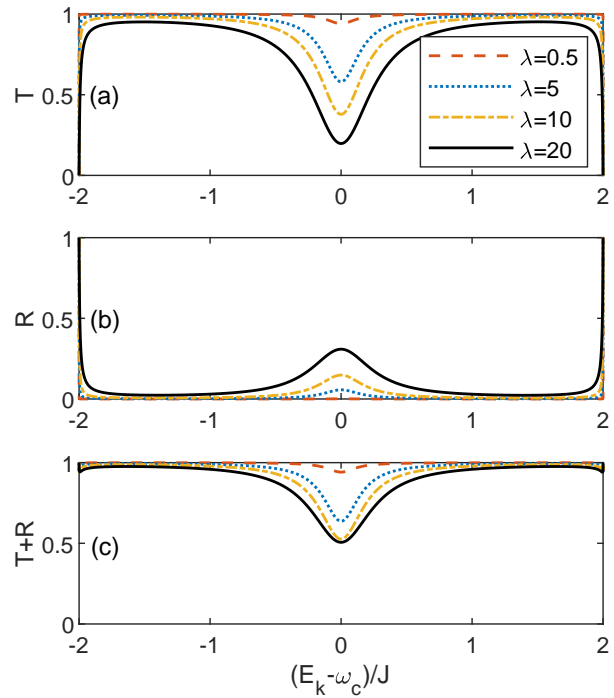}
	\caption{(color online) (a) The transmission coefficient, (b) the reflection coefficient {and (c) their summation} as a function of energy detuning $(E_k - \omega_{\mathrm{c}})$ at $\Omega=\omega_{\mathrm{c}}=10$, $J=4$, $\gamma=0.5$ for $\lambda=0.5$ (red dashed), $5$ (blue dotted), $10$ (yellow dash-dot) and $20$ (black solid).}
	\label{lambda}
\end{figure}

For the  Hermitian scattering center, such as the two-level system and perfect cavity \textit{et al.} \cite{PhysRevLett.101.100501,Zhou:20}, it usually satisfies $T+R=1$, which is a signature of probability current conservation. The incident photon can { either transmit through the scattering center or be reflected by it}. The non-Hermitian scattering center introduces loss and gain, which can break the probability current conservation, except for some $\mathcal{PT}$ symmetric systems \cite{PhysRevA.85.012111,PhysRevA.91.042131}. Although the imperfect cavity is a Hermitian scattering center, the probability current  may not be conserved  {either}. From Eqs. (\ref{T_k}) and (\ref{R_k}), one can easily obtain that
\begin{equation}
T+R=1+\frac{4J\sin k%
	\mathrm{Im}\left( V\right) }{\mathrm{Re}\left( V\right) ^{2}+\left( 2J\sin k-%
	\mathrm{Im}\left( V\right) \right) ^{2}} .
\label{T+R}
\end{equation}
Clearly, the imaginary part of the effective potential breaks the probability current conservation. This is understandable, since the imperfect cavity has infinite modes which can be regarded as a reservoir. The incident photon  { can not only be reflected  or transmitted, but also be kept in the imperfect cavity}.

To further understand the influence of the imperfect cavity on the scattering process, we firstly plot the transmission and reflection coefficients for different coupling strength $\lambda$, as shown in Fig. \ref{lambda}. The vanishing transmission coefficient at $E_k-\omega_{\mathrm{c}}=\pm 2 J$ correspond to $k=\pi$ and $0$ respectively. The coupling strength has significant influences on the scattering process, especially near the resonant frequency $E_k \simeq \omega_{\mathrm{c}}$. In the weak coupling regime, the imperfect cavity plays a relatively small role, which leads to the large transmission coefficient even at the resonant frequency $E_k=\omega_{\mathrm{c}}$. However, the total reflection always emerges at the resonant frequency for the perfect cavity. With the increase of the coupling strength $\lambda$, the transmission coefficient becomes smaller while the reflection coefficient becomes larger.  {Besides,  the summation of the reflection and transmission coefficients becomes smaller, as shown in Fig. \ref{lambda} (c). It indicates that the incident photon has an increasing probability to be kept in the imperfect cavity with the increase of the coupling strength $\lambda$, especially near the resonant frequency $E_k \simeq \omega_{\mathrm{c}}$.}

\begin{figure}[htbp]
	\centering
	\includegraphics[width=9cm]{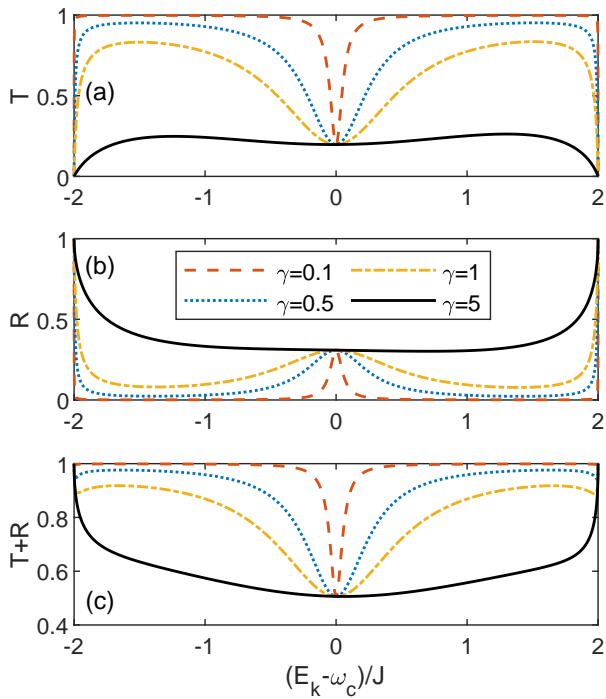}
	\caption{(color online) (a) The transmission coefficient, (b) the reflection coefficient {and (c) their summation} as a function of energy detuning $(E_k - \omega_{\mathrm{c}})$ at $\Omega=\omega_{\mathrm{c}}=10$, $J=4$, $\lambda=20$ for $\gamma=0.1$ (red dashed), $0.5$ (blue dotted), $1$ (yellow dash-dot) and $5$ (black solid).}
	\label{gamma}
\end{figure}
\begin{figure}[htbp]
	\centering
	\includegraphics[width=9cm]{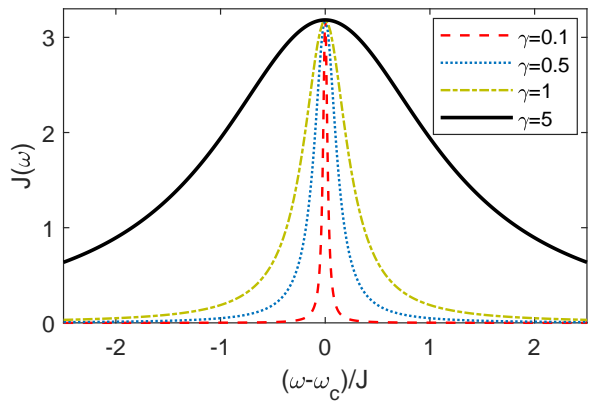}
	\caption{(color online) The spectral density $J(\omega)$ as a function of $\omega$ at $\omega_{\mathrm{c}}=10$, $\lambda=20$ for $\gamma=0.1$ (red dashed), $0.5$ (blue dotted), $1$ (yellow dash-dot) and $5$ (black solid).}
	\label{spectra}
\end{figure}
We then plot the transmission and reflection coefficients for different spectral width $\gamma$, as shown in Fig. \ref{gamma}. Under the off-resonant condition, the transmission  (reflection) coefficient is greatly suppressed (enhanced) with the increase of $\gamma$,  {while the summation of the reflection and transmission coefficients becomes much smaller}.
The off-resonant condition means that $|E_k-\omega_{\mathrm{c}}| \gg 0$. As we have mentioned above, the cavity mode with $\omega=E_k$ plays the dominant role, which is also far away from $\omega_{\mathrm{c}}$. The larger $\gamma$ leads to wider spectral $J(\omega)$, as shown in Fig. \ref{spectra}. It indicates that the coupling strength is large even if $\omega$ is far away from the $\omega_{\mathrm{c}}$. It is explicit that the reflection and transmission coefficients at the resonant frequency are robust against the change of $\gamma$. On one hand, we can attribute it to  { $J(\omega=\omega_{\mathrm{c}})$ which is independent of $\gamma$ }. On the other hand, we reexamine the effective potential $V$. At the resonant frequency $E_k=\omega_{\mathrm{c}}$, the effective potential can be simplified as
\begin{eqnarray}
V=\frac{\lambda}{2 \pi} \left(\ln \frac{\omega_{\mathrm{c}}}{\sqrt{\omega_{\mathrm{c}}^2 + \gamma^2}} - i \pi\right).
\end{eqnarray}
Clearly, when $\gamma$ is small enough,  $\mathrm{Re}(V)$ tends to zero. The effective potential is dominated by the imaginary part $\mathrm{Im}(V)$ which doesn't depend on $\gamma$. $\mathrm{Re}(V)$ comes into play only if $\gamma$ is very large.

\begin{figure}[htbp]
	\centering
	\includegraphics[width=9cm]{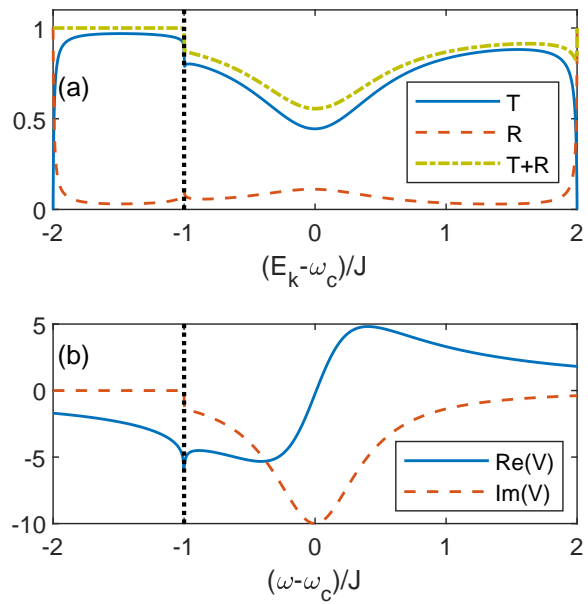}
	\caption{(color online) The transmission coefficient $T$ (blue solid line),  reflection coefficient $R$ (red dashed line)  {and their summation $T + R$ (yellow dash-dot line)} as a function of energy detuning $(E_k - \omega_{\mathrm{c}})$, (b) The real (blue solid line) and imaginary (red dash line) parts of the effective potential energy at $\Omega=\omega_{\mathrm{c}}=10$, $J=10$, $\lambda=20$, and $\gamma=4$.  {The vertical black line shows $E_k=0$ and $\omega=0$ respectively.}}
	\label{large_J}
\end{figure}

 {A perfect cavity coupled with the 1D CRW is a good candidate to be the quantum optical switch \cite{PhysRevLett.101.100501}. It can be either a perfect mirror at the resonant frequency which totally reflects photons, or an ideal transparent medium far away from the resonant frequency which allows photons to pass. However, no cavities are perfect due to the environment-induced dissipation in the experiments. Our studies on the imperfect cavity indicate that a large coupling strength $\lambda$ is necessary to maintain the high reflection coefficient at the resonant frequency as shown in Fig. \ref{lambda}, while a small spectral width $\gamma$ is necessary to maintain the high transmission coefficient away from the resonant frequency as shown in Fig. \ref{gamma}.}

The influence of the hopping amplitude $J$ on the transmission and reflection  also deserves some attention. The  range of $E_k \in \left[\Omega-2J,\Omega+2J\right]$ increases with the increase of $J$. One can expect that increasing $J$ can lead to larger detuning $|E_k - \omega_{\mathrm{c}}|$ which enhances the transmission but suppresses the reflection. One should also note that when $J>\Omega/2$, $E_k$ can be negative which leads to the disappear of the imaginary part of the effective potential (\ref{Im_V}).  {From Eq. (\ref{T+R}), it is clear that the the probability current conservation $T+R=1$ is preserved if there is no imaginary part of the effective potential at $E_k<0$. The incident photon with $E_k<0$ cannot resonate with any modes of imperfect cavity which suppresses its probability to be kept in the cavity.} When we continuously change the energy $E_k$ of the incident photon, the imaginary part of the effective potential jumps from zero when $E_k<0$ to a finite value when $E_k>0$, as shown in Fig. \ref{large_J} (b). The jumps of the effective potential leads to the discontinuity of  the transmission and reflection coefficients,  {as well as their summation,} as shown in Fig. \ref{large_J} (a).

\section{Conclusion}

We study the coherent transport of a single photon in a CRW. An imperfect cavity is coupled to the CRW which serves as the scattering center to control the transmission and the reflection of the incident photon. The dissipation is taken into account by assuming a Lorentzian spectrum for the cavity. In the weak dissipation limit, the imperfect cavity recovers the behavior of perfect one. For a finite dissipation, we find that the probability current  conservation can be broken, even if it is a Hermitian system. The incident photon cannot only be reflected or transmitted, but also be kept in the imperfect cavity.

The coupling strength $\gamma$ between the imperfect cavity and the CRW has significant influences on the scattering process, especially near the resonant frequency $E_k \simeq \omega_{\mathrm{c}}$. With the increase of the coupling strength, the transmission coefficient becomes smaller while the reflection coefficient becomes larger.  {Besides, their summation becomes smaller as the incident photon has more probability to be kept in the imperfect cavity.} The spectral width $\gamma$ plays a dominant role under the off-resonant condition, where the transmission  (reflection) coefficient is greatly suppressed (enhanced) with the increase of $\gamma$. It becomes very robust against the change of $\gamma$ at the resonant frequency.  {To make the  imperfect cavity a valid  quantum optical switch, one need to increase the coupling strength and decrease the spectral width.}

We also observe a discontinuity of the transmission and reflection coefficients. When the hopping amplitude $J$ of the CRW is large enough, the energy of the incident photon can be either positive or negative and the consequent scattering by the cavity varies greatly.  The imperfect cavity induces a complex effective potential, which is  closely related to all the observed phenomena.

\section*{ACKNOWLEDGEMENTS}
This work is supported by the National Science
Foundation of China (Grant Nos. 11834005 and  11674285).

\end{document}